\documentclass[preprint,showpacs,preprintnumbers]{revtex4}%
\usepackage{amsmath}
\usepackage{graphicx}
\usepackage{dcolumn}
\usepackage{bm}
\usepackage{hyperref}
\usepackage{amsmath}
\usepackage{amsfonts}
\usepackage{amssymb}%
\providecommand{\U}[1]{\protect\rule{.1in}{.1in}}
\begin{document}
\title{Relativistic theory of di-Holeums - quantized gravitational bound states of
two micro black holes}
\author{A. L. Chavda$^{a}$, L. K. Chavda$^{b}$}
\email{$^{a}$\ a01.l.chavda@gmail.com, $^{b}$\ holeum@gmail.com}
\affiliation{$^{a}$\ Physics Department, V.N. South Gujarat University, Udhna-Magdalla
Road, Surat - 395007, Gujarat, India. $^{b}$\ 49 Mahatma Gandhi Society, City
Light Road, Surat 395007, Gujarat, India.}
\date{\today }

\begin{abstract}
The Klein-Gordon equation is solved for di-Holeums (gravitational bound states
of two micro black holes) for scalar and vector gravity in its static limit.
The relativistic models confirm the predictions of the nonrelativistic
Newtonian gravity model, correct to about six significant figures over almost
the entire sub-Planck domain. All three models possess a mass range devoid of
physics. This is interpreted as evidence that the universe must have more than
four dimensions. We show that the formation of Holeums is feasible both in the
sub-Planck mass and above-Planck mass ranges.

\end{abstract}

\pacs{03.65.Ge, 03.65.Pm, 04.60.-m, 04.30.-w}
\maketitle
\tableofcontents

\section{Introduction}

The nature of dark matter is one of the most profound mysteries of our times.
Numerous candidates for dark matter particles have been proposed, which
include Standard Model neutrinos \cite{G.Bertone.et.al, L.Bergstrom}, Sterile
neutrinos \cite{G.Bertone.et.al, S.Dodelson.et.al}, WIMPs
\cite{R.Bernabei.et.al}, MACHOs \cite{The.MACHO.Collaboration}, Supersymmetric
particles \cite{S.P.Martin}, Kaluza-Klein matter \cite{H.Cheng.et.al}, Axions
\cite{Kawasaki.Nakayama}, Cryptons \cite{K.Benakli.et.al}, and primordial
black holes (PBHs) \cite{P.Frampton.et.al}, among others.

In 2002 we proposed another type of dark matter particle consisting of stable
atoms of PBHs, which we called the Holeum \cite{Chavda.Chavda.1}. Since then
we have shown that Holeums can be microscopic (such as a di-Holeum consisting
of two micro black holes) as well as macroscopic (such as a stellar-mass macro
Holeum consisting of $k$ micro black holes, where $k\gg2$), and can give rise
to quantized gravitational radiation \cite{Chavda.Chavda.2}. We have also
proposed that Holeums can give rise to cosmic rays of all observed energies,
including UHECR \cite{Chavda.Chavda.3}. Al Dallal has shown that the
observations on the very short duration Gamma Ray Bursts are well explained by
the Holeum model \cite{Dallal.1}. Al Dallal has also shown that the Holeum
model predicts diffuse Gamma Ray Halos around supernova remnants
\cite{Dallal.2}.

Exactly solvable problems have a special niche in theoretical physics. With
their help we can grasp the physics of a phenomenon much more clearly and
easily. Here we are trying to develop an effective model of quantum gravity
for the bound state problem of the di-Holeum. To this end, we investigated the
problem of a bound state of two micro black holes (MBHs) in quantized
Newtonian gravity using the non-relativistic Schr\"{o}dinger equation.and
obtained order of magnitude, ball-park values of the bound state parameters in
our 2002 paper \cite{Chavda.Chavda.1}. In this paper, we include special
relativity in the model. This is necessary because we are dealing with
ultra-compact MBHs and ultra-short distances. We solve the Klein-Gordon
equation for the di-Holeum, assuming gravitation to be either a scalar or a
vector interaction. Both of these cases are exactly solvable for a static
$1/r$ potential.

In \S \ II we present scalar gravity. We calculate the energy eigenvalues, the
mass, the binding energy and the radius of the ground state and the
characteristic function of the Holeum which determines whether the bound state
is a stable Holeum or an unstable one. In \S \ III we present similar
calculations for the case of vector gravity. In \S \ IV we present a summary
of corresponding results for the case of Newtonian gravity to facilitate their
comparison with the relativistic models. In \S \ V we compare the predictions
of these three models. Discussions and conclusions are presented in \S \ VI.
In the following treatment, as in \cite{Chavda.Chavda.1}, we use the term
"Holeum" to refer to the di-Holeum for the sake of simplicity.

\section{SCALAR GRAVITY}

\subsection{The Klein-Gordon Equation for Scalar Gravity}

The free particle relativistic Klein-Gordon equation is obtained from%
\begin{equation}
E^{2}=\mathbf{\vec{p}}^{2}c^{2}+\mu^{2}c^{4} \label{1}%
\end{equation}
by letting%
\begin{equation}
E\rightarrow i\hbar\frac{\partial}{\partial t},\mathbf{\vec{p}}\rightarrow
-i\hbar\mathbf{\vec{\nabla}} \label{2}%
\end{equation}
Here $E$ and $\mathbf{\vec{p}}$ are the total energy and the linear momentum
of a particle of reduced mass $\mu$, respectively and $\mathbf{\vec{\nabla}}$
is the three dimensional gradient operator. We are considering two identical
black holes of mass $m$ each and having no charge and spin. $\hbar$ and $c$
are the Planck's constant reduced by $2\pi$ and the speed of light in vacuum,
respectively. Thus, we get the free particle Klein-Gordon equation%
\begin{equation}
-\hbar^{2}\frac{\partial^{2}\psi}{\partial t^{2}}=\left(  -\hbar^{2}%
c^{2}\mathbf{\vec{\nabla}}^{2}+\mu^{2}c^{4}\right)  \psi\label{3}%
\end{equation}
Here $\psi$ is the wave function of the system. The static gravitational
interaction potential is given by%
\begin{equation}
V\left(  r\right)  =-\hbar c\frac{\alpha_{g}}{r} \label{4}%
\end{equation}
where%
\begin{equation}
\alpha_{g}=\left(  \frac{m}{m_{P}}\right)  ^{2} \label{5}%
\end{equation}
and%
\begin{equation}
m_{P}=\left(  \frac{\hbar c}{G}\right)  ^{\frac{1}{2}} \label{6}%
\end{equation}
Here $m_{P}$ is the Planck mass $1.22101\times10^{19}$\ GeV$/c^{2}$ and $G$ is
Newton's universal constant of gravity. $\alpha_{g}$ is the dimensionless
gravitational coupling constant. To treat $V\left(  r\right)  $, given by Eq.
(\ref{1}), as a scalar interaction we make the following substitutions in the
free-particle Klein-Gordon equation:%
\begin{align}
\mathbf{\vec{p}}  &  \rightarrow\mathbf{\vec{p}}\label{7a}\\
E  &  \rightarrow E\label{7b}\\
\mu c^{2}  &  \rightarrow\mu c^{2}+V\left(  r\right)  \label{7c}%
\end{align}
Thus we have%
\begin{equation}
-\hbar^{2}\frac{\partial^{2}\psi}{\partial t^{2}}=\left[  -\hbar^{2}%
c^{2}\mathbf{\vec{\nabla}}^{2}+\left(  \mu c^{2}+V\left(  r\right)  \right)
^{2}\right]  \psi\label{8}%
\end{equation}
This is the Klein-Gordon equation for scalar gravity. To obtain the stationary
state solution we let%
\begin{equation}
\psi(\mathbf{\vec{r}},t)=\psi\left(  \mathbf{\vec{r}}\right)  e^{\left(
\frac{-iEt}{\hbar}\right)  } \label{9}%
\end{equation}
Substituting this into Eq. \ref{8} we have%
\begin{equation}
E^{2}\psi=\left[  -\hbar^{2}c^{2}\mathbf{\vec{\nabla}}^{2}+\left(  \mu
c^{2}+V\left(  r\right)  \right)  ^{2}\right]  \psi\label{10}%
\end{equation}
We assume the separability of the wave function%
\begin{equation}
\psi\left(  \mathbf{\vec{r}}\right)  =\frac{R\left(  r\right)  }{r}P\left(
\theta\right)  Q\left(  \phi\right)  \label{11}%
\end{equation}
The solution for $Q$ is given by%
\begin{equation}
Q=e^{\left(  \text{im}\phi\right)  },\text{ }m=0,\pm1,\pm2,\ldots\label{12}%
\end{equation}
$P\left(  \theta\right)  $ satisfies the associated Legendre differential
equation%
\begin{equation}
\left(  P\sin\theta\right)  ^{-1}\left(  \frac{d}{d\theta}\right)  \left(
\sin\theta\frac{dP}{d\theta}\right)  +l\left(  l+1\right)  -\frac{m^{2}}%
{\sin^{2}\theta}=0 \label{13}%
\end{equation}
where%
\begin{equation}
l=0,1,2,\ldots\label{14}%
\end{equation}
The radial part $R\left(  r\right)  $ satisfies the equation%
\begin{equation}
\frac{R^{\prime\prime}\left(  r\right)  }{R\left(  r\right)  }+\left[
\frac{E^{2}-\left(  \mu c^{2}+V\left(  r\right)  \right)  ^{2}}{\hbar^{2}%
c^{2}}-\frac{l\left(  l+1\right)  }{r^{2}}\right]  =0 \label{15}%
\end{equation}
Substituting for $V\left(  r\right)  $ from Eq. \ref{4} we have%
\begin{equation}
R^{\prime\prime}\left(  r\right)  +\left[  \frac{E^{2}-\mu^{2}c^{4}}{\hbar
^{2}c^{2}}+\frac{\alpha_{g}}{\lambdabar r}+\frac{C}{r^{2}}\right]  R\left(
r\right)  =0 \label{16}%
\end{equation}
where%
\begin{equation}
\lambdabar=\frac{\hbar}{mc} \label{17}%
\end{equation}
and%
\begin{equation}
C=-\alpha_{g}^{2}-l\left(  l+1\right)  =\frac{1}{4}-q^{2} \label{18}%
\end{equation}
Let%
\begin{equation}
\kappa^{2}=\frac{\mu^{2}c^{4}-E^{2}}{\hbar^{2}c^{2}} \label{19a}%
\end{equation}%
\begin{equation}
B=\frac{\alpha_{g}}{\lambdabar} \label{20}%
\end{equation}%
\begin{equation}
q^{2}=\alpha_{g}^{2}+\left(  l+\frac{1}{2}\right)  ^{2} \label{21}%
\end{equation}
Substituting Eqs. \ref{18} to \ref{21} into Eq. \ref{16} we get%
\begin{equation}
R^{\prime\prime}\left(  r\right)  +\left[  -\kappa^{2}+\frac{B}{r}+\frac
{\frac{1}{4}-q^{2}}{r^{2}}\right]  R\left(  r\right)  =0 \label{22}%
\end{equation}
Define%
\begin{equation}
\rho=2\kappa r \label{23}%
\end{equation}
Substituting this into Eq. \ref{22} we have%
\begin{equation}
R^{\prime\prime}\left(  \rho\right)  +\left[  -\frac{1}{4}+\frac{\lambda_{0}%
}{\rho}+\frac{\frac{1}{4}-q^{2}}{\rho^{2}}\right]  R\left(  \rho\right)  =0
\label{24}%
\end{equation}
where%
\begin{equation}
\lambda_{0}=\frac{\mu c^{2}\alpha_{g}}{\left(  \mu^{2}c^{4}-E^{2}\right)
^{\frac{1}{2}}} \label{25}%
\end{equation}
Let%
\begin{equation}
R\left(  \rho\right)  =\rho^{\beta}e^{\left(  -\rho/2\right)  F\left(
\rho\right)  } \label{26}%
\end{equation}
This takes Eq. \ref{24} to the form%
\begin{equation}
\frac{F^{\prime\prime}\left(  \rho\right)  }{F\left(  \rho\right)  }+\left(
\frac{2\beta}{\rho}-1\right)  \frac{F^{\prime}\left(  \rho\right)  }{F\left(
\rho\right)  }+\frac{\left(  \lambda_{0}-\beta\right)  }{\rho}+\frac{\left(
\frac{1}{4}-q^{2}\right)  }{\rho^{2}}+\frac{\beta\left(  \beta-1\right)
}{\rho^{2}}=0 \label{27}%
\end{equation}
To cancel the last two terms let%
\begin{equation}
\frac{1}{4}-q^{2}=-\beta\left(  \beta-1\right)  =\frac{1}{4}-\left(
\beta-\frac{1}{2}\right)  ^{\frac{1}{2}} \label{28}%
\end{equation}%
\begin{equation}
\beta=\frac{1}{2}\pm q \label{29}%
\end{equation}
We take the positive square root to avoid the singularity at the origin in Eq.
\ref{26}. Thus we have%
\begin{equation}
\beta=\frac{1}{2}+\left[  \alpha_{g}^{2}+\left(  l+\frac{1}{2}\right)
^{2}\right]  ^{\frac{1}{2}} \label{30}%
\end{equation}
Then we can rewrite Eq. \ref{27} as%
\begin{equation}
\rho F^{\prime\prime}\left(  \rho\right)  +\left(  2\beta-\rho\right)
F^{\prime}\left(  \rho\right)  +\left(  \lambda_{0}-\beta\right)  F\left(
\rho\right)  =0 \label{31}%
\end{equation}
We compare this with the differential equation of the confluent hypergeometric
function given by%
\begin{equation}
xy^{\prime\prime}\left(  x\right)  +\left(  b-x\right)  y^{\prime}\left(
x\right)  -ay\left(  x\right)  =0 \label{32}%
\end{equation}
The general solution to this equation is given by%
\begin{equation}
y\left(  x\right)  =c^{\prime}{}_{1}F_{1}\left(  a,b;x\right)  +d^{\prime
}U\left(  a,b;x\right)  \label{33}%
\end{equation}
Here $c^{\prime}$ and $d^{\prime}$ are constants. The $U\left(  a,b;x\right)
$ is singular at $x=0$. Therefore we discard it by choosing $d^{\prime}=0$.
The other solution is the confluent hypergeometric function given by%
\begin{equation}
y={}_{1}F_{1}\left(  a,b;x\right)  =%
%TCIMACRO{\dsum \limits_{n=0}^{\infty}}%
%BeginExpansion
{\displaystyle\sum\limits_{n=0}^{\infty}}
%EndExpansion
\left(  a\right)  _{n}\frac{x^{n}}{\left(  b\right)  _{n}n!} \label{34}%
\end{equation}
where%
\begin{equation}
\left(  a\right)  _{n}=\frac{\Gamma\left(  a+n\right)  }{\Gamma\left(
a\right)  } \label{35}%
\end{equation}
This solution is nonsingular at the origin but it blows up at infinity. To
avoid this we require that it become a polynomial for large $x$. Then we see
from Eq. \ref{26} that it represents a localized wave function suitable for a
bound state. This is true only if%
\begin{equation}
a=-v,\text{ }v=0,1,2,...\infty\label{36}%
\end{equation}
Comparing Eqs. \ref{31} and \ref{32} we have%
\begin{align}
a  &  =\beta-\lambda_{0}\label{37a}\\
b  &  =2\beta\label{37b}%
\end{align}
From Eqs. \ref{36}, \ref{37a} and \ref{37b} we have%
\begin{equation}
\lambda_{0}=\beta+v=v+\frac{1}{2}+q \label{38}%
\end{equation}

\subsection{Energy Eigenvalues of a Holeum}

In the weak coupling limit $\alpha_{g}\ll1$, we have%
\begin{equation}
q=\left[  \alpha_{g}^{2}+\left(  l+\frac{1}{2}\right)  ^{2}\right]  ^{\frac
{1}{2}}\approx l+\frac{1}{2}+\frac{\alpha_{g}^{2}}{2l+1} \label{39}%
\end{equation}
Substituting this into Eq. \ref{38} we have%
\begin{equation}
\lambda_{0}=n+\frac{\alpha_{g}^{2}}{2l+1} \label{40}%
\end{equation}
where%
\begin{equation}
n=v+l+1=1,2,3,\ldots\label{41}%
\end{equation}
This is the principal quantum number of the bound state as we shall see below.
Substituting this into Eq. \ref{38}, we have%
\begin{equation}
\lambda_{0}=n+\epsilon_{l} \label{42}%
\end{equation}
where%
\begin{equation}
\epsilon_{l}=\left[  \alpha_{g}^{2}+\left(  l+\frac{1}{2}\right)  ^{2}\right]
^{\frac{1}{2}}-\left(  l+\frac{1}{2}\right)  \label{43}%
\end{equation}
Now from Eqs. \ref{25} and \ref{42} we have%
\begin{equation}
E_{nl}=\mu c^{2}\left[  1-\frac{\alpha_{g}^{2}}{\left(  n+\epsilon_{l}\right)
^{2}}\right]  ^{\frac{1}{2}} \label{44}%
\end{equation}
where we have shown only the positive sign of the square root for convenience.
Note that the negative sign, too, is equally admissible. With the advent of
the relativistic Klein-Gordon and the Dirac equations, there arose the need to
take the negative energy solutions seriously. As is well-known, the negative
energy solution was assigned to an antiparticle having the same mass as the
particle but having opposite quantum numbers. In our case, we have a Holeum
consisting of two PBHs. But an anti-Holeum consisting of two anti-PBHs is also
equally possible. This is because we are considering the very early universe
which was matter-antimatter symmetric before the decoupling of gravity. This
will lead to the well-known problem of explaining the absence of antimatter in
the present universe. We will not address it here. In the following we will
consider mainly the positive energy solutions. But we may also indicate the
results for the anti-Holeums. Now for $\alpha_{g}\ll1$ and $n\gg\epsilon_{l}$
we have%
\begin{equation}
E_{nl}=\mu c^{2}-\frac{\mu c^{2}\alpha_{g}^{2}}{2n^{2}} \label{45}%
\end{equation}
where we have kept only the first two terms in the expansion in powers of
$\alpha_{g}^{2}$. The first term is the rest energy and the second one is the
usual formula for the energy eigenvalues of the hydrogen atom. Therefore the
$n$ that occurs in Eq. \ref{45} and that is given by Eq. \ref{41} is
identified as the principal quantum number of the Holeum which is the
gravitational analogue of the hydrogen atom.

\subsection{Binding Energy and Mass of a Holeum}

The interaction energy is defined by%
\begin{equation}
W_{nl}=E_{nl}-\mu c^{2} \label{46}%
\end{equation}
The binding energy is given by%
\begin{equation}
B_{nl}=\left\vert W_{nl}\right\vert =\mu c^{2}\left\vert \left[
1-\frac{\alpha_{g}^{2}}{\left(  n+\epsilon_{l}\right)  ^{2}}\right]
^{\frac{1}{2}}-1\right\vert \label{47}%
\end{equation}
Note that the binding energy for the negative energy solution is also given by
the same expression.

For the $1s$ state the binding energy is given by%
\begin{equation}
B_{1s}=\frac{mc^{2}}{2}\left\vert 1-\left(  \frac{2}{p_{0}+1}\right)
^{\frac{1}{2}}\right\vert \label{48}%
\end{equation}%
\begin{equation}
p_{0}=\left(  1+4\alpha_{g}^{2}\right)  ^{\frac{1}{2}} \label{49}%
\end{equation}
For $\alpha_{g}\ll1$ the binding energy is given by%
\begin{equation}
B_{1s}=\left(  \frac{mc^{2}\alpha_{g}^{2}}{4}\right)  \left\vert
1-\frac{7\alpha_{g}^{2}}{4}\right\vert +o\left(  \alpha_{g}^{\frac{13}{2}%
}\right)  \label{50}%
\end{equation}
The mass of the Holeum is given by%
\begin{equation}
M_{n}=2m+\frac{W_{nl}}{c^{2}} \label{51}%
\end{equation}
Substituting from Eqs. \ref{44} and \ref{46} into Eq. \ref{51} we have%
\begin{equation}
M_{n}=\frac{m}{2}\left[  3+\left\{  1-\frac{\alpha_{g}^{2}}{\left(
n+\epsilon_{l}\right)  ^{2}}\right\}  ^{\frac{1}{2}}\right]  \label{52}%
\end{equation}
For the $1s$ state this is given by%
\begin{equation}
M_{1s}=\frac{m}{2}\left[  3+\left(  \frac{2}{p_{0}+1}\right)  ^{\frac{1}{2}%
}\right]  \label{53}%
\end{equation}
For $\alpha_{g}\ll1$ this is given by%
\begin{equation}
M_{1s}=2m\left(  1-\frac{\alpha_{g}^{2}}{8}\right)  +o\left(  \alpha
_{g}^{\frac{9}{2}}\right)  \label{54}%
\end{equation}
The first term on the right side of this equation is the same as the mass of a
Holeum in Newtonian gravity as we shall see later. Note that the last term in
this equation is less than $10^{-6}$ even for $m=10^{18}$ GeV$/c^{2}$.

\subsection{The Radius and the Characteristic Function of the Ground State}

The normalized radial part of the wave function is given by%
\begin{equation}
\psi\left(  r\right)  =\left[  \frac{8\kappa^{3}\Gamma\left(  n-l\right)
}{\Gamma\left(  n-l+2q\right)  \left(  2n-2l+2q-1\right)  }\right]  ^{\frac
{1}{2}}\rho^{q-\frac{1}{2}}e^{-\frac{\rho}{2}}L_{n^{\prime}}^{2q}\left(
\rho\right)  \label{55}%
\end{equation}
where $L_{n}^{m}\left(  x\right)  $ is an associated Laguerre polynomial and
$n^{\prime}=n-l-1$. The probability density is given by%
\begin{equation}
P=r^{2}\left\vert \psi\left(  r\right)  \right\vert ^{2} \label{56}%
\end{equation}
Substituting Eq. \ref{55} into Eq. \ref{56} we have%
\begin{equation}
P=\frac{2\kappa\Gamma\left(  n-l\right)  \rho^{2q+1}e^{-\rho}\left\{
L_{n^{\prime}}^{2q}\left(  \rho\right)  \right\}  ^{2}}{\Gamma\left(
n-l+2q\right)  \left(  2n-2l+2q-1\right)  } \label{57}%
\end{equation}
where $\kappa$ and $q$ are given by Eqs. \ref{19a} and \ref{39}, respectively.
From Eq. \ref{57} one can show that the most probable radius of the ground
state $n=1,l=0$ is given by%
\begin{equation}
r_{1s}=\left(  \frac{\lambdabar}{2\alpha_{g}}\right)  \left(  1+p_{0}\right)
^{2} \label{58}%
\end{equation}
where $\lambdabar$ is given by Eq. \ref{17}.The mass of the ground state of
Holeum is given by Eq. \ref{53} and the Schwarzschild radius of the ground
state is given by%
\begin{equation}
R_{1s}=\frac{2M_{1s}G}{c^{2}} \label{59}%
\end{equation}
The characteristic function of a Holeum is defined by%
\begin{equation}
f_{1s}=\frac{R_{1s}}{r_{1s}} \label{60}%
\end{equation}
One can show that%
\begin{equation}
f_{1s}=\left(  p_{0}-1\right)  \frac{3+\left(  \frac{2}{p_{0}+1}\right)
^{\frac{1}{2}}}{2\left(  p_{0}+1\right)  } \label{61}%
\end{equation}
If $f_{1s}\geq1$ then the Schwarzschild radius of the bound state is greater
than the physical radius of the bound state. In this case the Holeum is a
black hole. We call it a Black Holeum (BH). It will emit Hawking radiation and
evaporate away. But if $f_{1s}<1$, then the Holeum is not a black hole. It
will not emit Hawking radiation even though it contains two black holes. It is
a stable bound state, as stable as a hydrogen atom of which it is a
gravitational analogue. One can show numerically that all Holeums with
constituent masses $m<m_{c}(1s)$ are stable but those not satisfying this
condition are BHs which will evaporate away. $m_{c}(1s)$ is given by%
\begin{equation}
m_{c}(1s)=1.2722m_{P} \label{62}%
\end{equation}

\subsection{Asymptotics}

The asymptotic form of the associated Laguerre polynomial is given by%
\begin{equation}
L_{n}^{\alpha}(x)=\pi^{-\frac{1}{2}}e^{\frac{x}{2}}x^{-\frac{\alpha}{2}%
-\frac{1}{4}}n^{\frac{\alpha}{2}-\frac{1}{4}}\cos\left[  2\left(  nx\right)
^{\frac{1}{2}}-\pi\frac{\alpha}{2}-\frac{\pi}{4}\right]  +O\left(
n^{\frac{\alpha}{2}-\frac{3}{4}}\right)  \label{63}%
\end{equation}
where $n\gg1$. Substituting this into Eq. \ref{57} we have%
\begin{equation}
P=\left(  \frac{\alpha_{g}}{2\pi\lambdabar}\right)  \rho^{\frac{1}{2}%
}n^{2q-\frac{5}{2}}\cos^{2}\phi\label{64}%
\end{equation}%
\begin{equation}
\phi=2\left(  n\rho\right)  ^{\frac{1}{2}}-q\pi-\frac{\pi}{4} \label{65}%
\end{equation}
where we have taken $l=0$ for the s-states for simplicity. For $n\gg2q$ one
can show that%
\begin{equation}
\frac{d}{d\rho}lnP=-\left(  2\tan\phi\right)  \left(  \frac{n}{\rho}\right)
^{\frac{1}{2}} \label{66}%
\end{equation}
This vanishes for $\phi=k\pi,k=0,\pm1,\pm2,\ldots$. Now for $n\gg2q$, from Eq.
\ref{65} we have%
\begin{equation}
\phi=2\left(  n\rho\right)  ^{\frac{1}{2}}=k\pi\label{67}%
\end{equation}
This gives us%
\begin{equation}
\rho=\frac{k^{2}\pi^{2}}{4n} \label{68}%
\end{equation}
Now $\psi_{n}\left(  r\right)  $ can have $n$ maxima. Therefore we have $k\leq
n$. At the $k^{th}$ maximum the probability is given by%
\begin{equation}
P_{k}=\frac{k\alpha_{g}n^{2q-3}}{4\lambdabar} \label{69}%
\end{equation}
This rises linearly with $k$. These maxima overlap and the peak with the
highest probability has $k=n$. Taking this value of $k$ we have%
\begin{equation}
r_{n}=\frac{\rho}{2\kappa}=\frac{n^{2}\pi^{2}\lambdabar}{4\alpha_{g}}
\label{70}%
\end{equation}
where we have used%
\begin{equation}
\kappa=\frac{\alpha_{g}}{2n\lambdabar} \label{71}%
\end{equation}
We also have%
\begin{equation}
\lambdabar=\frac{R}{2\alpha_{g}} \label{72}%
\end{equation}
where $R$ is the Schwarzschild radius of the constituent PBH. Substituting
this into Eq. \ref{70} we have%
\begin{equation}
r_{n}=\frac{n^{2}\pi^{2}R}{8\alpha_{g}^{2}} \label{73}%
\end{equation}
This formula was first derived in the framework of Newtonian gravity. In the
next subsection we will derive the same formula for the case of vector gravity.

\subsection{Classes of Holeum}

From Eq. \ref{52}, with $n\gg1$, we have%
\begin{equation}
M_{n}=2m\left(  1-\frac{\alpha_{g}^{2}}{8n^{2}}\right)  \label{74}%
\end{equation}
From this we can calculate the Schwarzschild radius:%
\begin{equation}
R_{n}=\frac{2M_{n}G}{c^{2}} \label{75}%
\end{equation}
Substituting Eq. \ref{74} into Eq. \ref{75}, we have%
\begin{equation}
R_{n}=2R\left(  1-\frac{\alpha_{g}^{2}}{8n^{2}}\right)  \label{76}%
\end{equation}
The characteristic function is given by%
\begin{equation}
f_{ns}=\frac{R_{n}}{r_{n}}=\frac{16\alpha_{g}^{2}}{\pi^{2}n^{2}}\left(
1-\frac{\alpha_{g}^{2}}{8n^{2}}\right)  \label{77}%
\end{equation}
where we have used Eqs. \ref{73} and \ref{76}. For further analysis we need
the following identity \cite{Chavda.Chavda.4}:%
\begin{equation}
\frac{16X}{\pi^{2}}\left(  1-\frac{X}{8}\right)  =1-\frac{32X_{1}X_{2}}%
{\pi^{2}} \label{78}%
\end{equation}
where%
\begin{equation}
X=\frac{\alpha_{g}^{2}}{n^{2}} \label{79}%
\end{equation}%
\begin{equation}
X_{1}=\frac{X}{4}-1+\Delta\label{80a}%
\end{equation}%
\begin{equation}
X_{2}=\frac{X}{4}-1-\Delta\label{81}%
\end{equation}%
\begin{equation}
\Delta=\left(  \frac{1-\pi^{2}}{32}\right)  ^{\frac{1}{2}}=0.83161 \label{82}%
\end{equation}
Substituting Eqs. \ref{78} through \ref{81} into Eq. \ref{77} we have%
\begin{equation}
f_{ns}=1-\frac{32X_{1}X_{2}}{\pi^{2}} \label{83}%
\end{equation}
From Eq. \ref{83} it is clear that if $X_{1}X_{2}>0$, then $f_{1s}<1$. This
will lead to stable Holeums. This gives rise to two cases: $(i)$ $X_{1}<0$ and
$X_{2}<0$ which we denote as the mass range $b$. This leads to the inequality%
\begin{equation}
0<m<0.905929\text{ }m_{P} \label{84}%
\end{equation}
and $(ii)$ $X_{1}>0$ and $X_{2}>0$. We shall denote this as the mass range
$a$. This leads to the inequality%
\begin{equation}
1.645217\text{ }m_{P}<m<1.681793\text{ }m_{P} \label{85}%
\end{equation}
On the other hand if $X_{1}X_{2}<0$, then $f_{1s}>1$.This will lead to bound
states which are black holes. We have called them BHs above. These are
unstable and will evaporate away. Since the stable Holeums in the mass range
$a$ are more massive than the Plank mass we have called them Hyper Holeums
(HHs). Up to now this classification of Holeums follows that of the Newtonian
gravity case exactly. However, from Eq. \ref{52} it is clear that $M_{n}$
cannot be zero for real values of $\alpha_{g}$. Thus, the Lux Holeum with mass
$M_{n}=0$ \cite{Chavda.Chavda.4} is ruled out in relativistic scalar gravity.
From the discussion following Eq. \ref{61} it is seen that in the case of
relativistic scalar gravity for small values of $n$ there are only two mass
ranges corresponding to the H and BH cases whereas for large $n$ we have just
seen three mass ranges: H, BH, HH.

\subsection{Quantized Gravitational Radiation}

From the foregoing formulae we can show that when a Holeum undergoes an atomic
transition from a higher ($n_{2}$, $l_{2}$) state to a lower ($n_{1}$, $l_{1}%
$) state it emits gravitational radiation of frequency given by%
\begin{equation}
\nu=\frac{mc^{2}\alpha_{g}^{2}}{4h}\left[  \left(  \frac{1}{n_{1}^{2}}%
-\frac{1}{n_{2}^{2}}\right)  -2\alpha_{g}^{2}\left\{  \frac{1}{n_{1}%
^{3}\left(  2l_{1}+1\right)  }-\frac{1}{n_{2}^{3}\left(  2l_{2}+1\right)
}\right\}  \right]  \label{86}%
\end{equation}
where the first term gives the non-relativistic contribution and the second
one gives the relativistic contribution. Since $m$ varies up to $m_{c}$ this
will be a band spectrum. For $m$ in the range $10^{10}$ GeV$/c^{2}$ to
$10^{12}$ GeV$/c^{2}$ we can show that the Holeums have sizes in the atomic
and the nuclear domains, respectively. They emit gravitational radiation in
the low and medium frequency domains, respectively.

\section{VECTOR GRAVITY}

\subsection{The Klein-Gordon Equation with Vector Interaction}

The relativistic Klein-Gordon equation for a particle subject to a vector
interaction $A^{\mu}=(V,-\mathbf{\vec{A}})$ is obtained by making the
substitution $\mathbf{\vec{p}}\rightarrow\mathbf{\vec{p}}-\mathbf{\vec{A}}$,
$E\rightarrow E-V$ in the free particle relativistic Klein-Gordon equation%
\begin{equation}
E^{2}=\mathbf{\vec{p}}^{2}c^{2}+\mu^{2}c^{4} \label{87}%
\end{equation}
where $\mu$ is the reduced mass of the particle. $E$, $\mathbf{\vec{p}}$ and
$c$ have their meanings as in \S \ II. In the static limit $\mathbf{\vec{A}%
}\rightarrow0$ we have%
\begin{equation}
\left(  E-V\right)  ^{2}\psi\left(  \mathbf{\vec{r}}\right)  =\left(
\mathbf{\vec{p}}^{2}c^{2}+\mu^{2}c^{4}\right)  \psi\left(  \mathbf{\vec{r}%
}\right)  \label{88}%
\end{equation}
Here%
\begin{equation}
E\rightarrow i\hbar\frac{\partial}{\partial t},\mathbf{\vec{p}}\rightarrow
-i\hbar\mathbf{\vec{\nabla}} \label{89}%
\end{equation}
This leads to%
\begin{equation}
\left(  i\hbar\frac{\partial}{\partial t}-V\right)  ^{2}\psi\left(
\mathbf{\vec{r}}\right)  =\left(  \mu^{2}c^{4}-\hbar^{2}c^{2}\mathbf{\vec
{\nabla}}^{2}\right)  \psi\left(  \mathbf{\vec{r}}\right)  \label{90}%
\end{equation}
The stationary state equation is obtained by letting%
\begin{equation}
\psi\left(  \mathbf{\vec{r}},t\right)  =\psi\left(  \mathbf{\vec{r}}\right)
e^{-\frac{iEt}{\hbar}} \label{91}%
\end{equation}
Substituting Eq. \ref{91} into Eq. \ref{90} we have%
\begin{equation}
\left(  E-V\right)  ^{2}\psi\left(  \mathbf{\vec{r}}\right)  =\left(  \mu
^{2}c^{4}-\hbar^{2}c^{2}\mathbf{\vec{\nabla}}^{2}\right)  \psi\left(
\mathbf{\vec{r}}\right)  \label{92}%
\end{equation}
This is the relativistic Klein-Gordon equation with vector interaction.

\subsection{The Energy Eigenvalues}

Eq. \ref{92} may be rewritten as%
\begin{equation}
-\hbar^{2}c^{2}\mathbf{\vec{\nabla}}^{2}\psi\left(  \mathbf{\vec{r}}\right)
=\left[  \left(  E-V\right)  ^{2}-\mu^{2}c^{4}\right]  \psi\left(
\mathbf{\vec{r}}\right)  \label{93}%
\end{equation}
We assume the separability of the wave function%
\begin{equation}
\psi\left(  \mathbf{\vec{r}}\right)  =\frac{R\left(  r\right)  }{r}P\left(
\theta\right)  Q\left(  \phi\right)  \label{94}%
\end{equation}
As before we have%
\begin{equation}
Q=e^{\left(  \text{im}\phi\right)  },\text{ }m=0,\pm1,\pm2,\ldots\label{95}%
\end{equation}
and%
\begin{equation}
P\left(  \theta\right)  =P_{l}^{m}\left(  \theta\right)  ,\text{
}l=0,1,2,\ldots\label{96}%
\end{equation}
is the associated Legendre polynomial. It can be shown that the radial part of
the wave function satisfies%
\begin{equation}
\frac{R^{\prime\prime}\left(  r\right)  }{R\left(  r\right)  }+\left[
\frac{\left(  E-V\right)  ^{2}-\mu^{2}c^{4}}{\hbar^{2}c^{2}}-\frac{l\left(
l+1\right)  }{r^{2}}\right]  =0 \label{97}%
\end{equation}
The static Newtonian potential is given by%
\begin{equation}
V\left(  r\right)  =-\frac{\hbar c\alpha_{g}}{r} \label{98}%
\end{equation}
as before. Substituting Eq. \ref{98} into Eq. \ref{97} we have%
\begin{equation}
R^{\prime\prime}\left(  r\right)  +\left[  \frac{E^{2}-\mu^{2}c^{4}}{\hbar
^{2}c^{2}}+\frac{2\alpha_{g}E}{\hbar cr}+\frac{\alpha_{g}^{2}-l\left(
l+1\right)  }{r^{2}}\right]  R\left(  r\right)  =0 \label{99}%
\end{equation}
Let%

\begin{equation}
\kappa^{2}=\frac{\mu^{2}c^{4}-E^{2}}{\hbar^{2}c^{2}} \label{100a}%
\end{equation}%
\begin{equation}
\rho=2\kappa r \label{101}%
\end{equation}%
\begin{equation}
\lambda=\frac{\alpha_{g}E}{\hbar c\kappa} \label{102}%
\end{equation}
Substituting these into Eq. \ref{99} we have%
\begin{equation}
R^{\prime\prime}\left(  \rho\right)  +\left[  -\frac{1}{4}+\frac{\lambda}%
{\rho}+\frac{\alpha_{g}^{2}-l\left(  l+1\right)  }{\rho^{2}}\right]  R\left(
\rho\right)  =0 \label{103}%
\end{equation}
Let%
\begin{equation}
R\left(  \rho\right)  =\rho^{s+1}e^{-\frac{\rho}{2}}w\left(  \rho\right)
\label{104}%
\end{equation}
Substituting this into Eq. \ref{103} and dividing by $R$ we have%
\begin{equation}
\frac{w^{\prime\prime}}{w}+\frac{s\left(  s+1\right)  }{\rho^{2}}%
+\frac{\lambda-s-1}{\rho}+\left(  \frac{2s+2}{\rho}-1\right)  \frac{w^{\prime
}}{w}+\frac{\alpha_{g}^{2}-l\left(  l+1\right)  }{\rho^{2}}=0 \label{105}%
\end{equation}
Let us choose%
\begin{equation}
s\left(  s+1\right)  =l\left(  l+1\right)  -\alpha_{g}^{2} \label{106}%
\end{equation}
By completing the squares in $s$ and $l$ we find that the solution to this
equation is given by%
\begin{equation}
s=-\frac{1}{2}\pm\left[  \left(  l+\frac{1}{2}\right)  ^{2}-\alpha_{g}%
^{2}\right]  \label{107}%
\end{equation}
For a non-singular behavior at $\rho=0$ we take the positive sign on the right
hand side of Eq. \ref{107}. This simplifies Eq. \ref{105} as follows%
\begin{equation}
\rho w^{\prime\prime}+\left(  2s+2-\rho\right)  w^{\prime}+\left(
\lambda-s-1\right)  w=0 \label{108}%
\end{equation}
We compare it with Kummer's differential equation:%
\begin{equation}
xy^{\prime\prime}\left(  x\right)  +\left(  b-x\right)  y^{\prime}\left(
x\right)  -ay\left(  x\right)  =0 \label{109}%
\end{equation}
with%
\begin{equation}
a=s+1-\lambda,\text{ }b=2s+2 \label{110}%
\end{equation}
Kummer's differential equation has the general solution%
\begin{equation}
y\left(  x\right)  =c_{1}{}_{1}F_{1}\left(  a,b;x\right)  +c_{2}U\left(
a,b;x\right)  \label{111}%
\end{equation}
where $c_{1}$ and $c_{2}$ are constants. $U\left(  a,b;x\right)  $ blows up at
$x=0$. Therefore we choose $c_{2}=0$. Now $_{1}F_{1}\left(  a,b;x\right)  $
blows up at $x=\infty$ unless $a=-v$ where $v$ is a positive integer. In the
latter case $_{1}F_{1}\left(  a,b;x\right)  $ becomes a polynomial and
$\psi\left(  r\right)  $ gets a localized form fit for a bound state. Eq.
\ref{108} has a solution given by%
\begin{equation}
w\left(  \rho\right)  ={}_{1}F_{1}\left(  -\lambda+s+1,2s+2;\rho\right)
\label{112}%
\end{equation}%
\begin{equation}
v=\lambda-s-1=0,1,2,\ldots\infty\label{113}%
\end{equation}%
\begin{equation}
\lambda=s+1+v=l+v+1+\left[  \left(  l+\frac{1}{2}\right)  ^{2}-\alpha_{g}%
^{2}\right]  ^{\frac{1}{2}}-\left(  l+\frac{1}{2}\right)  \label{114}%
\end{equation}
We define the principal quantum number%
\begin{equation}
n=v+l+1,\text{ }v,l=0,1,2,\ldots\infty\label{115}%
\end{equation}
Substituting Eqs. \ref{114} and \ref{115} into Eq. \ref{112} we have%
\begin{equation}
w\left(  \rho\right)  ={}_{1}F_{1}\left(  -v,2s+2;\rho\right)  \label{116}%
\end{equation}
We may rewrite Eq. \ref{114} as%
\begin{equation}
\lambda=n+\epsilon_{l},\text{ }\epsilon_{l}=\left[  \left(  l+\frac{1}%
{2}\right)  ^{2}-\alpha_{g}^{2}\right]  ^{\frac{1}{2}}-\left(  l+\frac{1}%
{2}\right)  \label{117}%
\end{equation}
From Eqs. \ref{117} and \ref{102} we have%
\begin{equation}
\left(  \mu^{2}c^{4}-E^{2}\right)  ^{\frac{1}{2}}=\frac{\alpha_{g}%
E}{n+\epsilon_{l}} \label{118}%
\end{equation}
This may be rewritten as%
\begin{equation}
E_{nl}=\frac{\mu c^{2}}{\left[  1+\left(  \frac{\alpha_{g}}{n+\epsilon_{l}%
}\right)  ^{2}\right]  ^{\frac{1}{2}}} \label{119}%
\end{equation}
where we have taken only the positive square root for convenience. These are
the energy eigenvalues for vector gravity.

\subsection{Mass and Binding Energy of Holeum}

The interaction energy is given by%
\begin{equation}
W_{nl}=E_{nl}-\mu c^{2} \label{120}%
\end{equation}
The mass of a Holeum is given by%
\begin{equation}
M_{nl}=2m+\frac{W_{nl}}{c^{2}} \label{121}%
\end{equation}
Substituting from Eqs. \ref{119} and \ref{120} into Eq. \ref{121} we have%
\begin{equation}
M_{nl}=\frac{m}{2}\left[  3+\frac{1}{\left[  1+\left(  \frac{\alpha_{g}%
}{n+\epsilon_{l}}\right)  ^{2}\right]  ^{\frac{1}{2}}}\right]  \label{122}%
\end{equation}
In particular, the $n=1$, $l=0$ or the $1s$ state has the mass%
\begin{equation}
M_{1v}=\frac{m}{2}\left[  3+\left(  \frac{p_{1}+1}{2}\right)  ^{\frac{1}{2}%
}\right]  \label{123}%
\end{equation}
For $\alpha_{g}^{2}\ll1$ we may expand this in powers of $\alpha_{g}^{2}$ to
get%
\begin{equation}
M_{1v}=2m\left(  1-\frac{\alpha_{g}^{2}}{8}\right)  +o\left(  \alpha
_{g}^{\frac{9}{2}}\right)  \label{124}%
\end{equation}
In Eq. \ref{123} $p_{1}$ is given by%
\begin{equation}
p_{1}=\left(  1-4\alpha_{g}^{2}\right)  ^{\frac{1}{2}} \label{125}%
\end{equation}
The binding energy for the $1s$ state is given by%
\begin{equation}
B_{1v}=\frac{mc^{2}}{2}\left\vert 1-\left(  \frac{p_{1}+1}{2}\right)
^{\frac{1}{2}}\right\vert \label{126}%
\end{equation}
For $\alpha_{g}^{2}\ll1$ we may expand this in powers of $\alpha_{g}^{2}$ to
get%
\begin{equation}
B_{1v}=\frac{mc^{2}\alpha_{g}^{2}}{4}\left(  1+\frac{5\alpha_{g}^{2}}%
{4}\right)  +o\left(  \alpha_{g}^{\frac{13}{2}}\right)  \label{127}%
\end{equation}

\subsection{The Radius and the Characteristic Function of the Ground State}

The radius of the $1s$ state is given by%
\begin{equation}
r_{1v}=R_{P}\left(  \frac{p_{1}+1}{2\alpha_{g}}\right)  ^{\frac{3}{2}}
\label{128}%
\end{equation}
For $\alpha_{g}^{2}\ll1$ we may expand this in powers of $\alpha_{g}^{2}$ to
get%
\begin{equation}
r_{1v}=\frac{R_{P}}{\alpha_{g}^{\frac{3}{2}}}\left(  1-\frac{3\alpha_{g}^{2}%
}{2}-\frac{9\alpha_{g}^{4}}{8}\right)  +O\left(  \alpha_{g}^{\frac{9}{2}%
}\right)  \label{129}%
\end{equation}
In Eqs. \ref{128} and \ref{129} $R_{P}$ is the Schwarzschild radius of a
Planck-mass PBH, i.e.%
\begin{equation}
R_{P}=\frac{2m_{P}G}{c^{2}} \label{130}%
\end{equation}
The characteristic function of the Holeum is given by%
\begin{equation}
f_{1v}=\frac{R_{1v}}{r_{1v}}=4\alpha_{g}^{2}\left[  \frac{3+\left(
\frac{p_{1}+1}{2}\right)  ^{\frac{1}{2}}}{\left[  2\left(  p_{1}+1\right)
\right]  ^{\frac{3}{2}}}\right]  \label{131}%
\end{equation}
For $\alpha_{g}^{2}\ll1$ we may expand this in powers of $\alpha_{g}^{2}$ to
get%
\begin{equation}
f_{1v}=2\alpha_{g}^{2}\left(  1+\frac{11\alpha_{g}^{2}}{8}\right)  +o\left(
\alpha_{g}^{6}\right)  \label{132}%
\end{equation}
From Eq. \ref{125} it is clear that if%
\begin{equation}
\alpha_{g}\geq\frac{1}{2} \label{133}%
\end{equation}
there will be no bound states at all. For the $1s$ state, numerically it is
found that we will have stable Holeums if $0<m<0.70107$ $m_{P}$. But there are
only unstable BHs for $0.70107$ $m_{P}<m<0.7071$ $m_{P}$.

\subsection{Asymptotics}

The normalized radial wave function $\psi\left(  r\right)  =\frac{R\left(
r\right)  }{r}$ is given by%
\begin{equation}
\psi\left(  r\right)  =N\rho^{\frac{p-1}{2}}e^{-\frac{\rho}{2}}L_{n^{\prime}%
}^{p}\left(  \rho\right)  \label{134}%
\end{equation}
where%
\begin{equation}
p=2\left[  \left(  l+\frac{1}{2}\right)  ^{2}-\alpha_{g}^{2}\right]
^{\frac{1}{2}} \label{135}%
\end{equation}%
\begin{equation}
n^{\prime}=n-l-1 \label{136}%
\end{equation}%
\begin{equation}
N^{2}=\frac{8\kappa^{3}\Gamma\left(  n-l\right)  }{\Gamma\left(  n-l+p\right)
\left(  2n-2l+p-1\right)  } \label{137}%
\end{equation}%
\begin{equation}
\kappa=\frac{\alpha_{g}}{2\lambdabar\left[  \alpha_{g}^{2}+\left(
n+\epsilon_{l}\right)  ^{2}\right]  ^{\frac{1}{2}}} \label{138}%
\end{equation}%
\begin{equation}
\epsilon_{l}=\left[  \left(  l+\frac{1}{2}\right)  ^{2}-\alpha_{g}^{2}\right]
^{\frac{1}{2}}-\left(  l+\frac{1}{2}\right)  \label{139}%
\end{equation}%
\begin{equation}
\lambdabar=\frac{\hbar}{mc} \label{140}%
\end{equation}%
\begin{equation}
\rho=2\kappa r \label{141}%
\end{equation}
The probability density defined by%
\begin{equation}
P=R^{2} \label{142}%
\end{equation}
is given by%
\begin{equation}
P=\Theta\rho^{p+1}e^{-\rho}\left[  L_{n^{\prime}}^{p}\left(  \rho\right)
\right]  ^{2} \label{143}%
\end{equation}
where%
\begin{equation}
\Theta=\frac{2\kappa\Gamma\left(  n-1\right)  }{\Gamma\left(  n-l+p\right)
\left(  2n-2l+p-1\right)  } \label{144}%
\end{equation}
The asymptotic behavior of $L_{n^{\prime}}^{p}\left(  \rho\right)  $ for
$n^{\prime}\gg1$ is given by%
\begin{equation}
L_{n^{\prime}}^{p}\left(  \rho\right)  =\pi^{-\frac{1}{2}}\rho^{-\frac{p}%
{2}-\frac{1}{4}}e^{\frac{\rho}{2}}\left(  n^{\prime}\right)  ^{\frac{p}%
{2}-\frac{1}{4}}\cos\Phi\label{145}%
\end{equation}
where%
\begin{equation}
\Phi=2\left(  n^{\prime}\rho\right)  ^{\frac{1}{2}}-\frac{p\pi}{2}-\frac{\pi
}{4} \label{146}%
\end{equation}
Substituting Eqs. \ref{145} and \ref{146} into Eq. \ref{143} and assuming
$n^{\prime}\gg1$ we have%
\begin{equation}
P=\frac{\Theta}{\pi}\rho^{\frac{1}{2}}\left(  n^{\prime}\right)  ^{p-\frac
{1}{2}}\cos^{2}\Phi\label{147}%
\end{equation}
For $n^{\prime}\gg1$ the logarithmic derivative of $P$ is given by%
\begin{equation}
\frac{1}{P}\frac{dP}{d\rho}=-2\left(  \frac{n^{\prime}}{\rho}\right)
^{\frac{1}{2}}\tan\Phi\label{148}%
\end{equation}
This vanishes for%
\begin{equation}
\Phi=k\pi,\text{ }k=0,\pm1,\pm2,\ldots\label{149}%
\end{equation}
As argued earlier we have $k\leq n$. From Eq. \ref{146} for $n^{\prime}\gg1$
we have%
\begin{equation}
\Phi=2\left(  n\rho\right)  ^{\frac{1}{2}} \label{150}%
\end{equation}
where we have replaced $n^{\prime}$ by $n$ for sufficiently large $n$. From
the last two equations we have%
\begin{equation}
\rho_{\max}=\frac{k^{2}\pi^{2}}{4n} \label{151}%
\end{equation}
at the position of the maximum value of $P$. Substituting from Eqs. \ref{138}
and \ref{141} into Eq. \ref{151} we have%
\begin{equation}
r_{kn}=\frac{k^{2}\pi^{2}\lambdabar}{4\alpha_{g}} \label{152}%
\end{equation}
Here we have replaced $r$ by $r_{kn}$ at the position of a maximum. Now for
$n$ $\gg1$ we can show from Eq. \ref{138} that%
\begin{equation}
\kappa=\frac{\alpha_{g}}{2n\lambdabar} \label{153}%
\end{equation}
We can also show that%
\begin{equation}
\lambdabar=\frac{R}{2\alpha_{g}} \label{154}%
\end{equation}
Substituting Eq. \ref{154} into Eq. \ref{152} we get%
\begin{equation}
r_{kn}=\frac{n^{2}\pi^{2}R}{8\alpha_{g}^{2}} \label{155}%
\end{equation}
As we have seen above, for $n$ $\gg1$ there is an overlapping of peaks and the
largest peak with the highest probability occurs at $k=n$. Therefore for large
$n$ and with $k=n$, we have%
\begin{equation}
r_{n}=\frac{n^{2}\pi^{2}R}{8\alpha_{g}^{2}} \label{156}%
\end{equation}
This is the formula we had derived for the case of Newtonian gravity
\cite{Chavda.Chavda.1}. It is also true for the relativistic scalar gravity
case as we have already seen above. Thus, we have a universal,
model-independent, result for large $n$. For $n$ $\gg1$ Eq. \ref{122} reduces
to%
\begin{equation}
M_{n}=2m\left(  1-\frac{\alpha_{g}^{2}}{8n^{2}}\right)  \label{157}%
\end{equation}
This is the same as the corresponding equation for Newtonian gravity, see Eq.
\ref{161} below. On carrying out the analysis for the classification of
Holeums, given in \S \ II-F above, we can show that the BHs and the HH classes
are unphysical because for $\alpha_{g}>\frac{1}{2}$ the various quantities
such as mass, binding energy, radius etc. become purely imaginary. Thus we are
left with only the Holeums having masses in the range given by $0<m<0.7071$
$m_{P}$.

\subsection{Gravitational Radiation}

When a Holeum in an excited state with a principal quantum number $n_{2}$ and
energy eigenvalue $E_{2}$ makes a transition to lower state with the
corresponding quantities $E1$ and $n_{1}$, with $n_{2}-n_{1}=2$, it emits
quantized gravitational radiation of frequency given by%
\begin{equation}
\nu=\frac{E_{2}-E_{1}}{h} \label{159}%
\end{equation}
where $E_{1}$ and $E_{2}$ are given by Eq. \ref{119}.

\section{SUMMARY OF NEWTONIAN GRAVITY}

In this paper we have investigated scalar gravity and vector gravity in the
framework of the relativistic Klein-Gordon equation. Earlier we have
investigated Newtonian gravity in the framework of the nonrelativistic
Schrodinger equation \cite{Chavda.Chavda.1}. In order to compare these three
models and to arrive at useful conclusions we present below the corresponding
results of Newtonian gravity:

If we solve the Schrodinger equation with the potential given in Eq.
\ref{4},we get the following energy eigenvalues:%
\begin{equation}
E_{n}=-\frac{\mu c^{2}\alpha_{g}^{2}}{2n^{2}} \label{160}%
\end{equation}
The mass of a Holeum is given by%
\begin{equation}
M_{n}=2m\left(  1-\frac{\alpha_{g}^{2}}{8n^{2}}\right)  \label{161}%
\end{equation}
The binding energy is given by%
\begin{equation}
B_{n}=\frac{mc^{2}\alpha_{g}^{2}}{4n^{2}} \label{162}%
\end{equation}
The most probable radius of the ground state is given by%
\begin{equation}
r_{1}=\frac{R}{\alpha_{g}^{2}} \label{163}%
\end{equation}
where%
\begin{equation}
R=\frac{2mG}{c^{2}} \label{164}%
\end{equation}
For $n$ $\gg1$ the most probable radius is given by%
\begin{equation}
r_{n}=\frac{\pi^{2}n^{2}R}{8\alpha_{g}^{2}} \label{165}%
\end{equation}
From Eq. \ref{159} we can get an expression for the Schwarzschild radius of
the $n^{th}$ excited state of a Holeum:%
\begin{equation}
R_{n}=2R\left(  1-\frac{\alpha_{g}^{2}}{8n^{2}}\right)  \label{166}%
\end{equation}
where%
\begin{equation}
R_{n}=\frac{2M_{n}G}{c^{2}} \label{167}%
\end{equation}

where $M_{n}$ is given by Eq. \ref{161}. From Eqs. \ref{165} and \ref{166} one
can calculate the characteristic function $f_{nN}$ for the Newtonian gravity
case. It is identical to that for the scalar gravity case given in \S \ II,
starting with Eq. \ref{78} and ending with Eq. \ref{83}. In other words, in
the asymptotic range $n$ $\gg1$ the results of the relativistic scalar case
are identical to those of the Newtonian gravity case.

\section{COMPARISION OF THE MODELS}

\subsection{Concordance between Newtonian gravity, relativistic scalar
gravity, and relativistic vector gravity}

In Table 1 we present the predictions of the binding energy of a Holeum by
three models, namely Newtonian gravity, relativistic scalar gravity, and
relativistic vector gravity. A study of the table reveals an interesting
concordance: Three different mathematical functions representing the binding
energy of a Holeum in three models under consideration give the same numerical
value of the binding energy correct to about six significant figures even as
the independent variable $m$ varies over seventeen orders of magnitude.
Similar statements can be made about the mass and other properties of the
Holeum predicted by the three models. This must be one of the rarest
concordances in theoretical physics. Note, however, that the predictions of
the relativistic models begin to disagree with those of the Newtonian gravity
case above $10^{17}$ GeV$/c^{2}$.

\begin{center}%
\begin{tabular}
[c]{|l|l|l|l|}\hline
$m$ GeV$/c^{2}$ & Relativistic Scalar Gravity & Relativistic Vector Gravity &
Newtonian Gravity\\\hline
$1\times10^{10}$ & $1.12476844\times10^{-27}$ & $1.12476844\times10^{-27}$ &
$1.12476844\times10^{-27}$\\\hline
$1\times10^{11}$ & $1.12476844\times10^{-22}$ & $1.12476844\times10^{-22}$ &
$1.12476844\times10^{-22}$\\\hline
$1\times10^{12}$ & $1.12476844\times10^{-17}$ & $1.12476844\times10^{-17}$ &
$1.12476844\times10^{-17}$\\\hline
$1\times10^{13}$ & $1.12476844\times10^{-12}$ & $1.12476844\times10^{-12}$ &
$1.12476844\times10^{-12}$\\\hline
$1\times10^{14}$ & $1.12476844\times10^{-7}$ & $1.12476844\times10^{-7}$ &
$1.12476844\times10^{-7}$\\\hline
$1\times10^{15}$ & $1.12476844\times10^{-2}$ & $1.12476844\times10^{-2}$ &
$1.12476844\times10^{-2}$\\\hline
$1\times10^{16}$ & $1.12476844\times10^{3}$ & $1.12476844\times10^{3}$ &
$1.12476844\times10^{3}$\\\hline
$1\times10^{17}$ & $1.12476844\times10^{8}$ & $1.12476844\times10^{8}$ &
$1.12476844\times10^{8}$\\\hline
$1\times10^{18}$ & $1.12467990\times10^{13}$ & $1.12483171\times10^{13}$ &
$1.12476844\times10^{13}$\\\hline
$1\times10^{19}$ & $6.75175221\times10^{17}$ & --- & $1.12476844\times10^{18}%
$\\\hline
$1.5\times10^{19}$ & $2.31197\times10^{18}$ & --- & $8.54121\times10^{18}%
$\\\hline
\end{tabular}

\bigskip\textbf{Table 1:} Ground state binding energy in $GeV$ of a Holeum in
relativistic scalar gravity, relativistic vector gravity and Newtonian gravity
\end{center}

\subsection{No physics up to $10^{10}$ GeV$/c^{2}$, followed by a mass range
containing interesting physics}

A remarkable fact that emerges from the Table 1 is that there is absolutely no
physics up to the constituent mass $10^{10}$ GeV$/c^{2}$. We emphasize that
this is the common feature of all the three models studied here. This is true
not only for the binding energy but also for the other properties of Holeums.
This is a "great desert". This is is followed by a mass range containing many
potentially very interesting physics phenomena: As shown in
\cite{Dallal.Azzam} and \cite{Chavda.Chavda.4} the Holeums of constituent
masses between $10^{10}$ GeV$/c^{2}$ and $10^{11}$ GeV$/c^{2}$ have roughly
atomic dimensions and they emit quantized gravitational radiation of low
frequencies in the kHz range accessible to LIGO (Laser Interferometer
Gravity-wave Observatory) and other gravity wave detectors. Holeums of masses
between $10^{11}$ GeV$/c^{2}$ and $10^{12}$ GeV$/c^{2}$ have roughly nuclear
dimensions. They emit gravitational waves of higher frequencies. Holeums of
masses up to $10^{14}$ GeV$/c^{2}$ have low binding energies up to about $112$
eV. In the Holeum model, Holeums of all masses accumulate overwhelmingly in
galactic halos as cold dark matter \cite{Chavda.Chavda.4}. They are prone to
break up due to collisions, which results in the emission of Hawking radiation
whose two components we have previously identified with cosmic rays and
gamma-ray bursts \cite{Chavda.Chavda.4}.

\subsection{Contrast between relativistic scalar gravity and relativistic
vector gravity}

In contrast to relativistic scalar gravity, relativistic vector gravity
imposes a sharp cut-off on the mass of the PBHs that can form stable Holeums.
This follows from the factor $\left(  1-4\alpha_{g}^{2}\right)  ^{\frac{1}{2}%
}$ appearing in the expressions for the energy eigenvalues, mass, binding
energy and the radius of the Holeum. This restricts Holeum formation to
strictly sub-Planck $m<0.70107$ $m_{P}$. The unstable BHs occur in the mass
range $0.70107$ $m_{P}<m<0.707107$ $m_{P}$.This is for the ground state. For
the highly excited states $n$ $\gg1$ the stable Holeum formation is restricted
to $m<0.707107$ $m_{P}$ and there are no BHs and HHs.

\section{DISCUSSIONS AND CONCLUSIONS}

In this paper we have studied the consequences of the Holeum conjecture in the
relativistic domain. We have solved the relativistic Klein-Gordon equation in
the limit of static gravity treating the latter either as a scalar or a vector
interaction. \emph{Our most interesting finding is that the relativistic
models confirm the predictions of the non-relativistic Newtonian gravity model
within about six significant figures over almost the entire sub-Planck domain.
This strongly validates our earlier results based on Newtonian gravity.}

The comparison of the two relativistic models with Newtonian gravity reveals
two very striking results: (1) As a function of the mass $m$ of the
constituent PBHs there is a great gap in the formation of viable Holeums as
reflected in the values of the parameters such as the binding energy. This gap
stretches over ten orders of magnitude in $m$. (2) This gap is followed by an
interval stretching over four orders of magnitude in $m$ containing
potentially very interesting physics such as quantized gravitational
radiation, cosmic rays, gamma ray bursts, etc.

The presence of such a great gap in a physical theory is spurious and
unnatural. It points to an inadequacy in the model. Since gravity and the
geometry of space-time are intrinsically inseparable \emph{the presence of the
gap in the three and the four dimensional models strongly suggests that our
universe must have more than four dimensions.} Since our model is exclusively
a model of gravity this finding has a more direct relevance to the
dimensionality of the space-time of the universe.

The presence of a \textquotedblleft great desert\textquotedblright\ is
well-known in particle physics. The electroweak unification in particle
physics occurs at 300 GeV but the next unification, called the Grand
Unification of the strong and the electro-weak interactions, occurs at
$10^{16}$ GeV. This is the "great desert" of particle physics. A remedy is
suggested by Kaluza-Klein theories with two extra dimensions
\cite{Arkani-Hamed.et.al}. They bring down the Planck energy scale to one TeV
when the two extra dimensions are compactified. This removes the
\textquotedblleft great desert\textquotedblright\ in particle physics.

Kaluza-Klein theory may have a two-fold effect on the Holeum model: (a) It may
remove our gap. (b) In such theories it is found that the dark matter
particles such as PBHs and Holeums are confined to a manifold different from
the one in which the particles of the standard model are confined
\cite{Dallal.Azzam}. This may have interesting implications for Holeum
formation in the domain of interesting physics.

Now let us address the issue of the verification of the Holeum conjecture. We
first list the criteria for the formation of stable, non-radiating bound
states of black holes, namely, Holeums. These are: (1) The rate of the
gravitational interactions $\Gamma$ must be greater than $H$, the rate of the
expansion of the universe at the temperature of the formation of the bound
state. (2) The binding energy of the would-be bound state must be much greater
than the kinetic energy of the primordial brew at the temperature under
consideration.(3) The ratio, $f_{n}$ of the Schwarzschild radius of the bound
state, $R_{n}$ to the most probable radius, $r_{n}$ of the bound state must be
less than unity. That is, $f_{n}<1$.

The first criterion simply says that the primordial brew is instantaneously in
thermal equilibrium. That is, it expands adiabatically. The first criterion is
a rule of the thumb. It is found to give surprisingly good results in general,
even though one must use the Boltzmann equation for more precise results in
actual practice. The second criterion ensures that the dissociation due to the
collisions is negligible. The third one says that the bound state of two black
holes is not itself a black hole. Otherwise it will be destroyed by the
Hawking radiation it must emit.

Now it is well-known \cite{Kolb.Turner} that perturbative interactions
mediated by massless gauge bosons such as the photons and the gravitons are
incapable of thermalizing the universe above the temperature of $10^{16}$ GeV.
But they can thermalize the universe below the latter temperature. That is,
the condition (1), $\Gamma>H$, is satisfied in the temperature range less than
$10^{16}$ GeV as mentioned above. For the Relativistic Scalar Gravity case, we
see from Eq. \ref{62} that the condition $f_{1s}<1$ is satisfied for
$m<1.2722$ $m_{P}=1.5521\times10^{19}$ GeV$/c^{2}$. This complies with the
condition (3). From Table 1 we see that the binding energies of the Holeums
with constituent mass in the range%
\begin{equation}
10^{19}\text{ GeV}/c^{2}<m<1.55\times10^{19}\text{ GeV}/c^{2} \label{168}%
\end{equation}

satisfy the condition (2) reasonably well. Since these Holeums have masses
greater than the Planck mass we call them the Hyper Holeums (HH)
\cite{Chavda.Chavda.4}. Thus the formation of HHs is feasible as in
\cite{Chavda.Chavda.4}.Note that the mass range in \cite{Chavda.Chavda.4} is
somewhat different from the scalar relativistic case.

Now let us consider the domain of interesting physics lying between $10^{10}$
GeV$/c^{2}$ and $10^{14}$ GeV$/c^{2}$. In particular consider the case
$m=10^{14}$ GeV$/c^{2}$ at a temperature of, say, $1$ eV. As noted above the
interactions mediated by massless gauge bosons, such as the graviton in this
case, are capable of thermalizing the universe below the temperature of
$10^{16}$ GeV. Thus the condition (1) mentioned above is satisfied. The
binding energy of a Holeum made of two black holes each of mass $10^{14}$
GeV$/c^{2}$ is about $112$ eV which is much greater than the kinetic energy at
the temperature of $1$ eV. Therefore at a temperature of the order of $1$ eV
the condition (2) on the binding energy is satisfied quite well. From Eq.
\ref{62} we see that the condition (3) $f_{1s}<1$ is also satisfied in the
case of the scalar gravity under consideration here. Thus, we conclude that
the formation of Holeums of mass$10^{14}$ GeV$/c^{2}$ is quite feasible at the
temperature of the order of $1$ eV or less. It is clear that a good case for
the formation of Holeums in the domain of interesting physics is at hand.

\end{document}